# Spectroscopic Fingerprint of Phase-Incoherent Superconductivity in the Cuprate Pseudogap State

Jhinhwan Lee[1], K. Fujita[1,2], A. R. Schmidt[1], Chung Koo Kim[1], H. Eisaki[3], S. Uchida[2], & J.C. Davis[1,4]

[1] *Laboratory of Atomic and Solid State Physics, Department of Physics, Cornell University, Ithaca, NY 14853, USA.*

[2] *Department of Physics, University of Tokyo, Bunkyo-ku, Tokyo 113-0033, Japan.*

[3] *Institute of Advanced Industrial Science and Technology, Tsukuba, Ibaraki 305-8568, Japan.*

[4] *Condensed Matter Physics and Material Science Department, Brookhaven National Laboratory, Upton, NY 11973, USA.*

**A possible explanation for the existence of the cuprate "pseudogap" state is that it is a d-wave superconductor without quantum phase rigidity. Transport and thermodynamic studies provide compelling evidence that supports this proposal, but few spectroscopic explorations of it have been made. One spectroscopic signature of d-wave superconductivity is the particle-hole symmetric "octet" of dispersive Bogoliubov quasiparticle interference modulations. Here we report on this octet's evolution from low temperatures to well into the underdoped pseudogap regime. No pronounced changes occur in the octet phenomenology at the superconductor's critical temperature $T_c$, and it survives up to at least temperature $T \sim 1.5T_c$. In the pseudogap regime, we observe the detailed phenomenology that was theoretically predicted for quasiparticle interference in a phase-incoherent d-wave superconductor. Thus, our results not only provide spectroscopic**

evidence to confirm and extend the transport and thermodynamics studies, but they also open the way for spectroscopic explorations of phase fluctuation rates, their effects on the Fermi arc, and the fundamental source of the phase fluctuations that suppress superconductivity in underdoped cuprates.

Superconductivity in the hole-doped $CuO_2$ Mott insulators is unique by virtue of the convergence of three phenomena. First, the superconductivity is quasi—two-dimensional, because it occurs primarily in the $CuO_2$ planes. Second, presumably because of correlations, the superfluid density is very low and increases from zero approximately linearly with the hole density $p$. Third, because it is a d-wave system, the superconducting energy gap $\Delta(\vec{k})$ exhibits four $k$-space nodes where zero-energy excitations exist. This unique combination means that fluctuations of the quantum phase $\phi(\vec{r},t)$, where $\vec{r}$ is the position vector and $t$ is time, of the superconducting order parameter $\Psi = \Delta e^{i\phi(\vec{r},t)}$ should have profound effects on cuprate superconductivity at low hole density (*1-6*). One might then expect to observe a sequence of three different temperature scales. The highest characteristic temperature would occur at the mean-field scale, $T_{MF}$, where local pairing begins. The next, $T_\phi$, is where strong diamagnetic fluctuations would set in. The lowest is the true superconducting critical temperature, $T_c$, where phase rigidity would appear in the presence of whatever combination of thermal and quantum fluctuations (*7-12*) actually exists in underdoped cuprates.

A schematic of the hole-doped cuprate phase diagram is shown in fig. S1A; within the gray shaded pseudogap regime, the approximate region where transport and thermodynamic evidence for phase-incoherent superconductivity has been detected (*13-18*) is shaded light blue. The techniques used include terahertz transport studies (*13*), the Nernst effect in thermal transport (*14,15*), torque-magnetometry measurements of diamagnetism (*16*), field dependence of the diamagnetism (*17*), and zero-bias conductance enhancement in tunnel junctions linking pseudogapped to superconducting samples (*18*). But the regions designated as containing phase-incoherent superconductivity differ markedly between these studies. Moreover, it has been proposed recently that much of the Nernst signature attributed to phase fluctuations may be due to the appearance of stripes (*19*). Thus, a direct spectroscopic fingerprint of phase-incoherent d-wave superconductivity could help to identify the precise regions dominated by phase fluctuations. Perhaps more importantly, the associated spectroscopic phenomena might be used to quantify phase fluctuation rates and to discriminate between different quantum fluctuation sources (*7-12*) causing the loss of phase rigidity, and thus suppression of high temperature superconductivity, in underdoped cuprates.

Spectroscopic imaging scanning tunneling microscopy (SI-STM) provides one approach to search for a spectroscopic fingerprint of phase-incoherent d-wave superconductivity. With this technique, the local density of states $N(\vec{r},E)$ (where $E$ is energy), the two branches of the Bogoliubov excitation spectrum $\vec{k}_B(\pm E)$, and the superconducting energy gap magnitude $\pm|\Delta(\vec{k})|$ for both filled and empty states can be determined in a single experiment (*20-23*). This is because a characteristic form of quasiparticle interference (QPI) occurs in cuprate superconductors where the Bogoliubov quasiparticle dispersion $E(\vec{k})$ has banana-shaped constant energy contours (*20-23*), as shown schematically in fig. S1B. In theory, the $k$-space locations of the Bogoliubov band minima and maxima, $\vec{k}_B(\pm E)$, coincide with the maxima in the joint density-of-states at the eight tips of these "bananas". Elastic scattering between these eight regions $\vec{k}_j(\pm E)$ (*j*=1,2,…,8) produces real-space interference modulations in $N(\vec{r},E)$ that are characterized by two sets of seven dispersive wavevectors $\vec{q}_i(\pm E)$ ($i$=1,2,...,7), where $\vec{q}_i(+E) = \vec{q}_i(-E)$ (fig. S1B). This is referred to as the octet model of cuprate QPI (*20-23*) and, for $T<<T_c$, it is a definitive signature of d-wave superconductivity. When these $\vec{q}_i(E)$ are measured from the Fourier transform of modulations in the differential tunneling conductance maps, $g(\vec{r},E) \equiv dI/dV(\vec{r},E)$, where $I$ is the tunnel current and $V$ is the tunnel-junction bias voltage, the Bogoliubov bands $\vec{k}_B(\pm E)$ and the superconductor's energy gap structure $\pm|\Delta(\vec{k})|$ can be determined (*20-23*). This technique is unique in that the measured gap $\pm|\Delta(\vec{k})|$ is definitely that of the superconductor and, because the technique uses visualization of interference patterns, it automatically identifies $k$-space regions having coherent quasiparticles.

A possible signature of phase-incoherent d-wave superconductivity in the pseudogap regime could be the continued existence of this QPI octet phenomenology. This is because, if the quantum phase $\phi(\vec{r},t)$ of $\Psi = \Delta e^{i\phi(\vec{r},t)}$ is fluctuating in space and time, then the energy gap magnitude $\pm|\Delta(\vec{k})|$ could still remain largely unchanged so that the particle-hole—symmetric octet of high joint—density-of-states regions generating the QPI would continue to exist. Because an ungapped Fermi arc appears above $T_c$ in underdoped cuprates (*24*), it is only the remaining gapped regions beyond the tips of the arc that would be available to support any such pseudogap QPI. Indeed, detailed theoretical studies of the QPI phenomenology that would be expected if the pseudogap regime is a phase-incoherent d-wave superconductor bear out this simple picture (*25-27*). Among their predictions is the existence of a particle-hole—symmetric octet of dispersive QPI modulations, but perhaps the octet might emerge from a $k$-space region that is different from that in the superconducting phase.

However, as revealed by the pioneering SI-STM studies done at temperatures above $T_c$ (28), the detection of the complete octet of dispersive QPI modulations in the pseudogap regime presents severe technical challenges. Because the basic observables are the particle-hole—symmetric dispersions $\vec{q}_i(+E) = \vec{q}_i(-E)$ for the seven wave vectors $\vec{q}_i$, the primary objectives are to achieve high precision in simultaneous measurements of all seven $|\vec{q}_i|$ and of $E$ within the pseudogap regime. Because the uncertainty in energy of a tunneling electron grows rapidly with temperature (exceeding 30 meV at 100 K), precision in the measurement of $E$ requires the lowest possible temperatures. We therefore chose to reduce $T_c$ of our $Bi_2Sr_2Ca_{0.8}Dy_{0.2}Cu_2O_{8+\delta}$ sample to 37±3 K by strong underdoping so that the pseudogap regime could be entered at low temperatures (fig. S1A, inset). Second, adequate $|\vec{q}_i|$ precision for $Bi_2Sr_2CaCu_2O_{8+\delta}$ QPI requires fields of view (FOVs) exceeding 45×45 nm$^2$ and any smaller FOV would unavoidably create the erroneous impression of nondispersive modulations (21). Thus, all studies herein were made on an FOV exceeding 45×45 nm$^2$ (for example, fig. S2). Even more challenging (as we show below) is that the intensities for some $\vec{q}_i(E)$ modulations in the pseudogap regime diminish by more than a factor of 10 compared with the superconducting state, so that greatly increased signal-to-noise ratios for all $g(\vec{r}, E)$ measurements are required. To achieve this increased ratio, we acquired each $g(\vec{r}, E)$ map for up to 10 days by using a specially designed low-noise preamplifier, a highly drift- and temperature-stabilized STM system and a replicating single-atom tip preparation technique. Finally, because the truly nondispersive electronic structure of the high-energy states (23, 29) is projected onto the low-energy $g(\vec{r}, E)$ by the systematic junction formation error (22, 23), the resulting spurious appearance of nondispersive signals at low bias is avoided here by the use of the relation $Z(\vec{r}, E) \equiv g(\vec{r}, +E)/g(\vec{r}, -E)$ for all QPI analyses (*22*). Using $Z(\vec{r},E)$ and $Z(\vec{q},E)$ also has the advantage that it automatically extracts particle-hole symmetric $\vec{q}_i(+E) = \vec{q}_i(-E)$ QPI modulations (*22,23*).

By combining all these capabilities in a specially designed, variable-temperature SI-STM system, we were able to study the temperature evolution of the octet of QPI modulations in $Z(\vec{q},E)$ from the superconducting phase into the pseudogap regime. The octet of interference modulations was studied at temperatures $T$ = 4.5, 15, 30, 37, 45, and 55 K, thereby entering well into the strongly underdoped pseudogap regime (fig. S1A). We measured the $g(\vec{r},E)$ images with subatomic resolution (fig. S2) on a single $Bi_2Sr_2Ca_{0.8}Dy_{0.2}Cu_2O_{8+\delta}$ sample with $p$ = 7±1 %. This data set consists of ~$5 \times 10^5$ atomically resolved and registered tunneling spectra and is described in the supporting online material (SOM) for every temperature *T,* in terms of a comprehensive set of $Z(\vec{r},E)$ images (fig. S3) and the related $Z(\vec{q},E)$ movies (movies S1 to S6).

Representative $Z(\vec{q},E)$ for six temperatures are shown in Fig. 1. Several important observations can be made from these data. First, the set of $\vec{q}_i(E)$ ($i$=1,2,…,7) that is characteristic of the superconducting octet model is preserved unchanged upon passing above $T_c$ and exists up to at least $T \sim 1.5T_c$ [compare, for example, (A) and (U)]. This result demonstrates that the QPI octet phenomenology exists in the pseudogap regime. Second, the existence of such interference patterns indicates that coherent wavelike quantum states occur in some parts of $k$-space in the pseudogap regime. Third, the octet of QPI wave vectors are quantitatively different at different temperatures, indicating that they are generated by different regions of $k$-space, and thus that $\Delta(\vec{k})$ is evolving with temperature. Fourth, some, but not all, QPI modulation intensities become far weaker in the pseudogap regime.

The measured values of $|\vec{q}_1(E)|$, $|\vec{q}_5(E)|$, and $|\vec{q}_7(E)|$ at each temperature shown in Fig. 2 reveal that the modulations that are dispersive in the superconducting phase (*20-23*) remain dispersive into the pseudogap regime. This is equally true for every measured octet wavevector (see fig. S4 and movies S1 to S6). Moreover, if d-wave energy gaps without particle-hole symmetry were generating these pseudogap QPI effects, instead of the 16 pairs of modulation wave vectors observed here, $Z(\vec{q}, E)$ should exhibit 32 pairs because then, $\vec{q}_i(+E) \neq \vec{q}_i(-E)$. These observations are important for discrimination between any static electronic ordered state having nondispersive modulations at an ordering wavevector $\vec{Q}$ and the dispersive k-space eigenstates of a phase-incoherent d-wave superconductor. Because we find that for |E| < 35 meV, all detected $\vec{q}_i(E)$ in the pseudogap regime are dispersive and particle-hole symmetric (Fig. 2 and fig. S4) and, moreover, that the dispersions are internally consistent with the octet model (fig. S4), we conclude that these low-energy modulations in $Z(\vec{q}, E)$ do not represent the signature of a static, electronic ordered state. However, the true nondispersive excitations of underdoped cuprates (*23,29*), which occur at the pseudogap energy scale (near $E \sim \pm 120$ meV in this sample) and break translational and rotational symmetry locally, remain completely unaltered upon the transition through $T_c$ into the pseudogap regime (fig. S7).

By requiring octet-model internal consistency (*20-23*) in the dispersions of all measured $\vec{q}_i(E)$ for all energies and all temperatures (fig. S4), we can determine $\vec{k}_B(\pm E)$ and $|\Delta(\vec{k})|$ as a function of *T*. The results shown in Fig. 3A (and figs. S5 and S6) indicate that, even at the lowest temperatures, the locus of scattering $\vec{k}_B(E)$ exhibits the ungapped arc previously reported in (*22,23,30*). Beyond the tips of this arc are the gapped regions that exhibit particle-hole—symmetric QPI, which then disappears at the line connecting ($\pi/a_0$,0) and (0,$\pi/a_0$) (*23*). Beyond that line, the electronic excitations to the pseudogap energy scale exhibit locally broken translational and rotational symmetries in $r$-space (*23,29*). Figure 3A shows the energy gap $|\Delta(\vec{k})|$ measured using the standard octet analysis (*20-23*) as a function of temperature. We see that, whereas the measured $\vec{k}_B(E)$ are on the same contour in $k$-space at all $T$ (fig. S5), the ungapped arc length grows slowly with increasing $T$ and the gapped regions exhibiting octet QPI thereby become more constricted in $k$-space (fig. S6). Figure 3B shows that the ungapped arc length is a monotonically increasing function of temperature (figs. S6 and S8). Moreover, the simultaneously measured zero-bias conductance shown in Fig. 3B (which is proportional to the density of states at the Fermi energy shown in Fig. 3C) also exhibits a high value at lowest temperatures and increases linearly with $T$. This result provides additional evidence for an ungapped zero-temperature arc that lengthens linearly with $T$. If such an ungapped arc in the superconducting state can be generated either by scattering (*31,32*) or by nonthermally generated phase fluctuations (*6*), then these results appear to be consistent with the linear increase in Fermi arc length with temperature that was originally observed in the pseudogap regime at higher doping (*33*).

We have emphasized that little distinguishes the observed octet QPI phenomenology between the superconducting phase and the underdoped pseudogap region (Figs. 1 to 3). However, Fig. 4 shows analysis of the energy and temperature dependence of the amplitudes of modulations due to scattering between $k$-space regions that have the same sign of the d-wave order parameter and between regions of opposite sign. Whereas the latter intensities drop precipitously as $T$ rises (Fig. 4, B to D), the former pass right through $T_c$ without appreciable changes (Fig. 4, F to H). This could be consistent with the d-wave Bogoliubov quasiparticle scattering processes, whose coherence factor products are insensitive to order parameter sign changes (*30*), remaining unperturbed by the loss of long-range phase rigidity in the pseudogap regime.

With the high $|\vec{q}|$ and $E$ resolution and the enhanced sensitivity to very weak conductance variations introduced here, we were able to find the following characteristics for the low-energy quasiparticle interference modulations in the underdoped cuprate pseudogap regime: (i) The set of seven $\vec{q}_i(E)$ ($i$=1,2,…,7) that are characteristic of the superconducting octet model are preserved unchanged upon passing above $T_c$ (Fig. 1 and fig. S4). (ii) All the $\vec{q}_i(E)$ remain dispersive in a manner that is internally consistent with the octet phenomenology (Fig. 2 and fig. S4). (iii) The octet modulation wave vectors retain their particle-hole symmetry $\vec{q}_i(+E) = \vec{q}_i(-E)$. (iv) The modulations occur in the same energy range and emanate from the same contour in $k$-space (although at varying locations on that contour) as those observed at lowest temperatures (Fig. 3 and fig. S5). (v) The particle-hole—symmetric energy gap $\pm|\Delta(\vec{k})|$ moves away from the nodes with increasing $T$, leaving behind a growing arc of gapless excitations (Fig. 3). (vi) The intensities of modulations due to scattering between $k$-space regions having the same sign of a d-wave order parameter are maintained, whereas the intensities of those having a different sign are greatly diminished (Fig. 4).

These observations represent a number of advances in our understanding of the electronic structure of the cuprate pseudogap regime. First, because the basic QPI octet phenomenology is definitely the signature of d-wave superconductivity well below $T_c$ (*20-23*), and because we show here that it evolves without pronounced changes through $T_c$, it seems implausible that it represents a different state above $T_c$. Moreover, the observed particle-hole—symmetric, dispersive, octet-model phenomenology is consistent with theoretical predictions for the QPI characteristics of a phase-incoherent d-wave superconductor (*25-27*). Our data therefore provide spectroscopic evidence confirming and extending the deductions reached in the transport and thermodynamic studies (*13-17*). In addition, knowledge of this spectroscopic signature may now make it possible to explore the phase diagram regions exhibiting this state, to quantify the phase fluctuation rates with changing temperature (*6*) and doping, and to discriminate between different quantum fluctuation sources (*7-12*). Second, because all the $\vec{q}_i(E)$ observed in $Z(\vec{q}, E)$ disperse internally consistently with the octet model, they do not represent the signature of a static, ordered state in the underdoped pseudogap regime. Nevertheless, it is equally important to emphasize that the true nondispersive and symmetry-breaking excitations that occur at the pseudogap energy scale (*23,29*) coexist with these low-energy dispersive modulations at all temperatures studied. Third, analysis of the data using the octet model provides a new perspective on the superconducting energy gap at lowest dopings, indicating that an arc of gapless excitations exists in the strongly underdoped superconducting phase (*22, 23, 30*) and that it expands linearly in temperature. Such an arc might exist either due to effects of scattering (*31,32*) or due to phase fluctuations (*6*) that are generated by a purely quantum mechanical processes (*7-12,34*), or both. Fourth, as the alterations in interference intensities (Fig. 4) that are detected with increasing temperature mimic those generated by introduction of static vortices at low temperatures (*30*), it may be that they occur here for similar reasons but now in a nonstatic vortex fluid (*13-17*). Finally, our observations reveal that the electronic structure of the strongly underdoped pseudogap regime contains not two, but three fundamental components: (i) the Fermi arc (*24*), (ii) the gap $\pm|\Delta(\vec{k})|$ of a phase-disordered superconductor to coherent *k*-space excitations, and (iii) the nondispersive and locally symmetry-breaking excitations at the pseudogap energy scale (*23,29*).

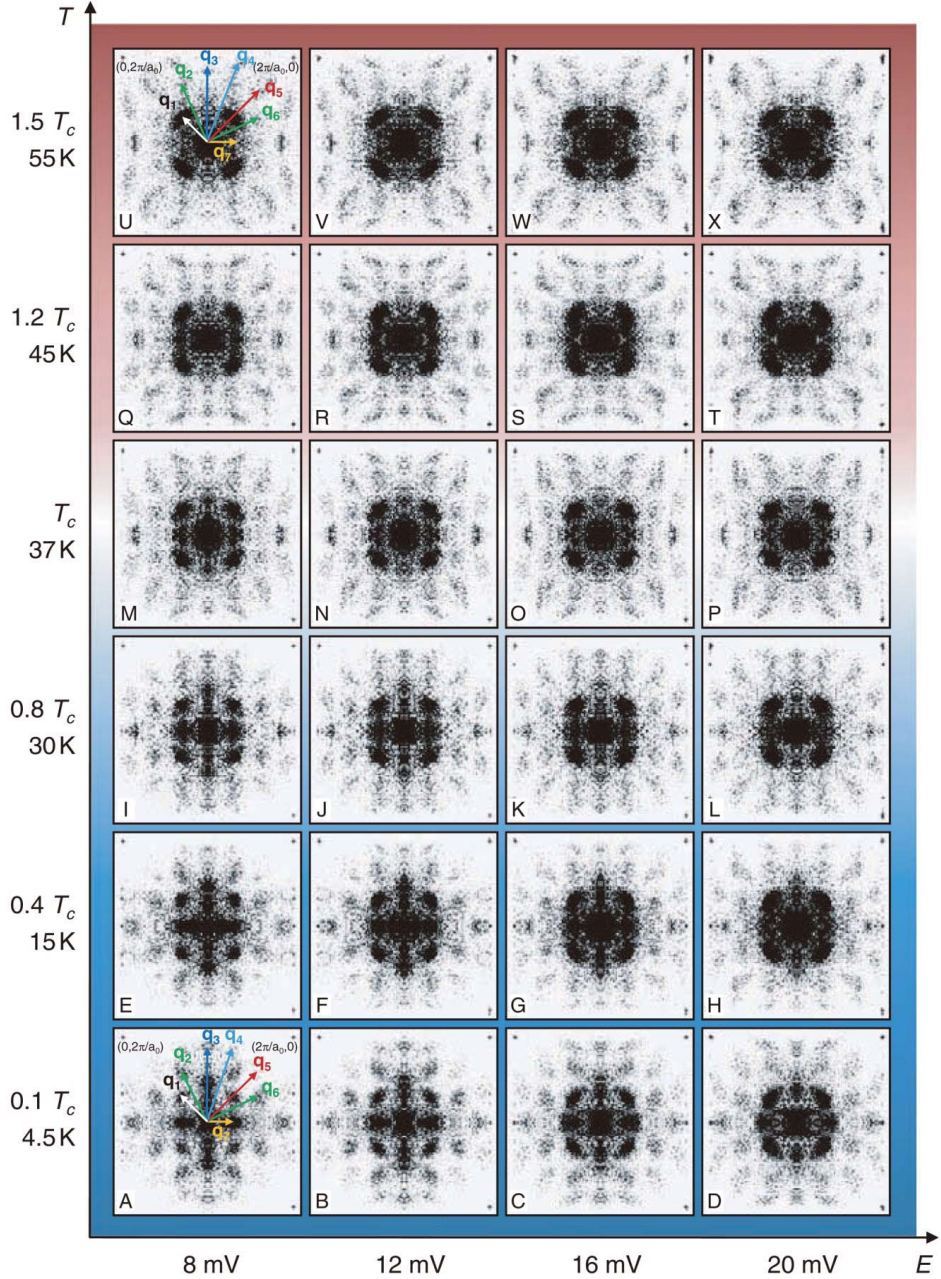

**Figure 1.** (**A** to **X**) Differential conductance map $g(\vec{r}, E) = dI/dV(\vec{r}, E = eV)$ were obtained on the same sample in an atomically resolved and registered FOV > 45×45 nm$^2$ at six temperatures (fig. S2). Each panel shown is the Fourier transform $Z(\vec{q}, E)$ of $Z(\vec{r}, E) \equiv g(\vec{r}, +E)/g(\vec{r}, -E)$ for a given energy and temperature (fig. S3). The QPI signals evolve dispersively with energy along the horizontal energy axis. The temperature dependence of QPI for a given energy evolves along the vertical axis. The octet-model set of QPI wave vectors is observed for every $E$ and $T$ (see movies S1 to S6) as seen, for example, by comparing (A) and (U), each of which has the labeled octet vectors. Within the basic octet QPI phenomenology, there is no particular indication in these data of where the superconducting transport $T_c$ occurs.

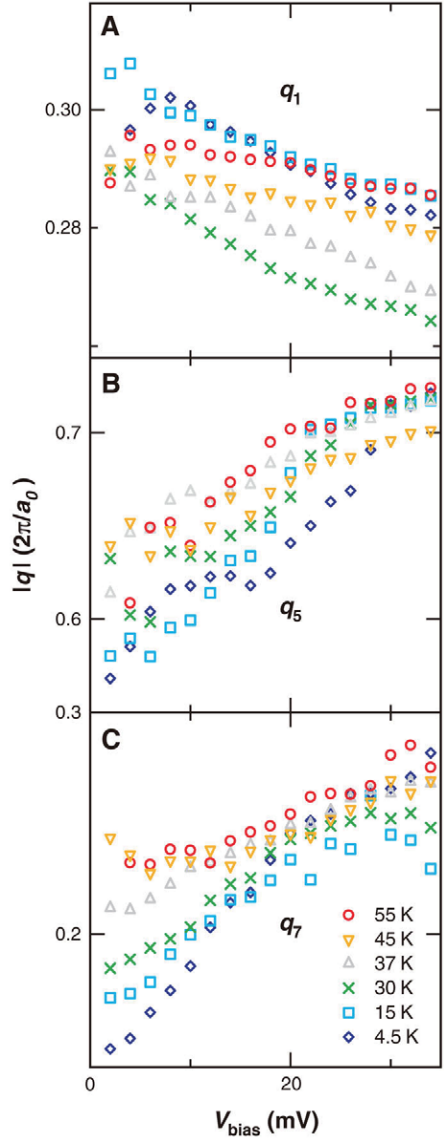

**Figure 2.** (**A** to **C**) Temperature dependence of the length of representative $q$-vectors— $|\vec{q}_1(E)|$, $|\vec{q}_5(E)|$, and $|\vec{q}_7(E)|$ — as a function of energy. We show here and in fig. S4 and movies S1 to S6 that octet modulation vectors $\vec{q}_1$, $\vec{q}_2$, $\vec{q}_4$, $\vec{q}_5$, $\vec{q}_6$, and $\vec{q}_7$ are manifestly dispersive at all temperatures studied, with the gradual changes in length as temperature is increased (fig. S3). Wave vector $\vec{q}_3$ has been difficult to observe at all energies and so is not used in the QPI analysis. Wave vector $\vec{q}_5$ is easily detected and is dispersive as expected; however, it contains a second-order contribution and so is not used in the determination of $\pm|\Delta(\vec{k})|$.

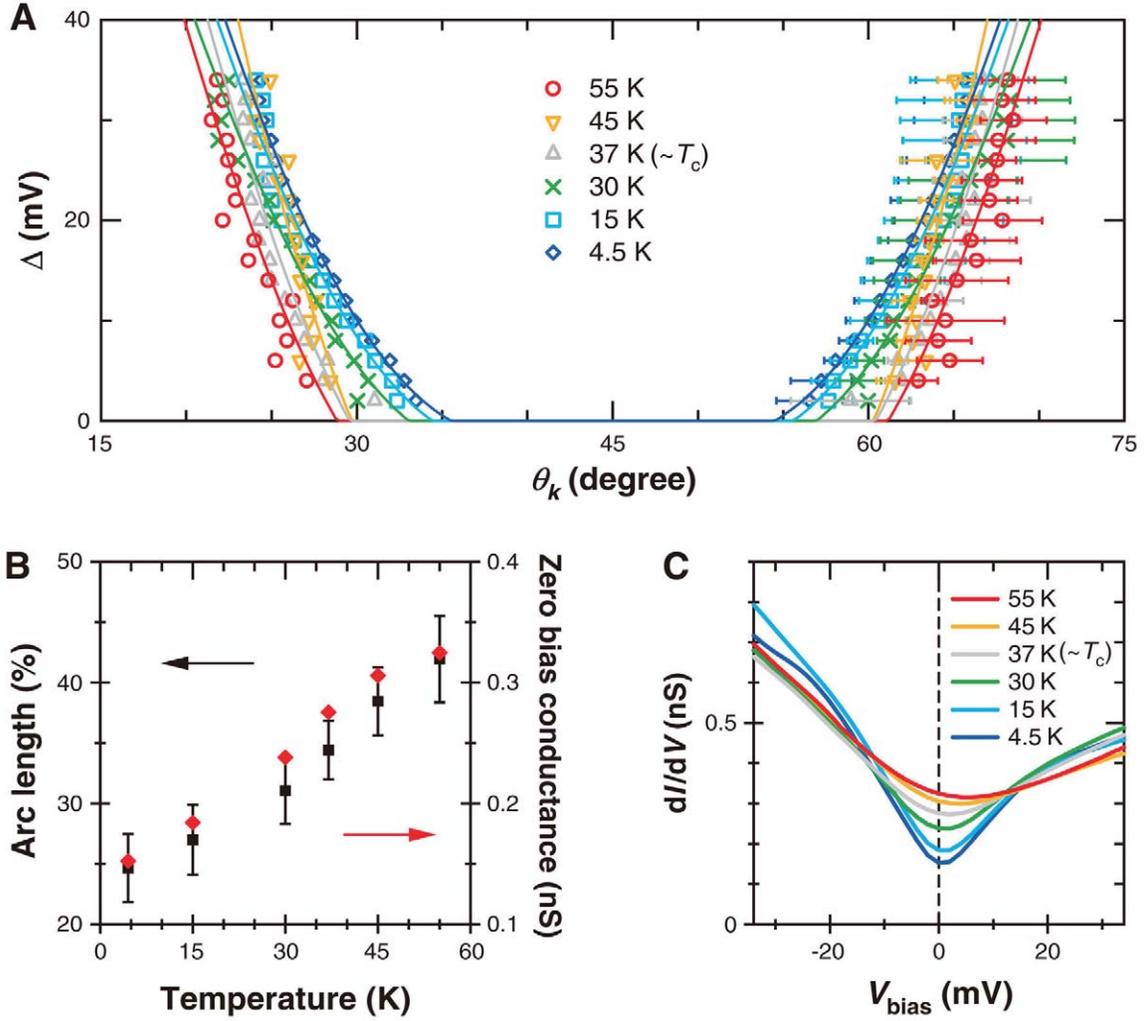

**Figure 3.** (**A**) The superconducting energy gap $\pm|\Delta(\vec{k})|$ to coherent $k$-space Bogoliubov quasiparticle excitations [as determined from the internally consistent octet model analysis of $Z(\vec{q}, E)$] in the superconducting phase is shown for temperatures 4.5, 15 and 30 K. The energy gap magnitude $\pm|\Delta(\vec{k})|$ to coherent $k$-space excitations in the nonsuperconducting phase or pseudogap regime is shown for temperatures 37, 45, and 55 K. (**B**) The ungapped arc length from QPI analysis shows approximately linear temperature dependence. As previously reported in (*22*, *23*, *30*), there is an ungapped region even at the lowest temperatures in the superconducting state and, based on the observations shown here, this is probably indicative of combined effects of disorder and phase fluctuations from quantum mechanical sources. (**C**) The spatially averaged differential conductance is plotted as a function of temperature measured simultaneously with the data shown in Fig. 1. The high value of the zero-bias conductance is apparent even at low temperatures.

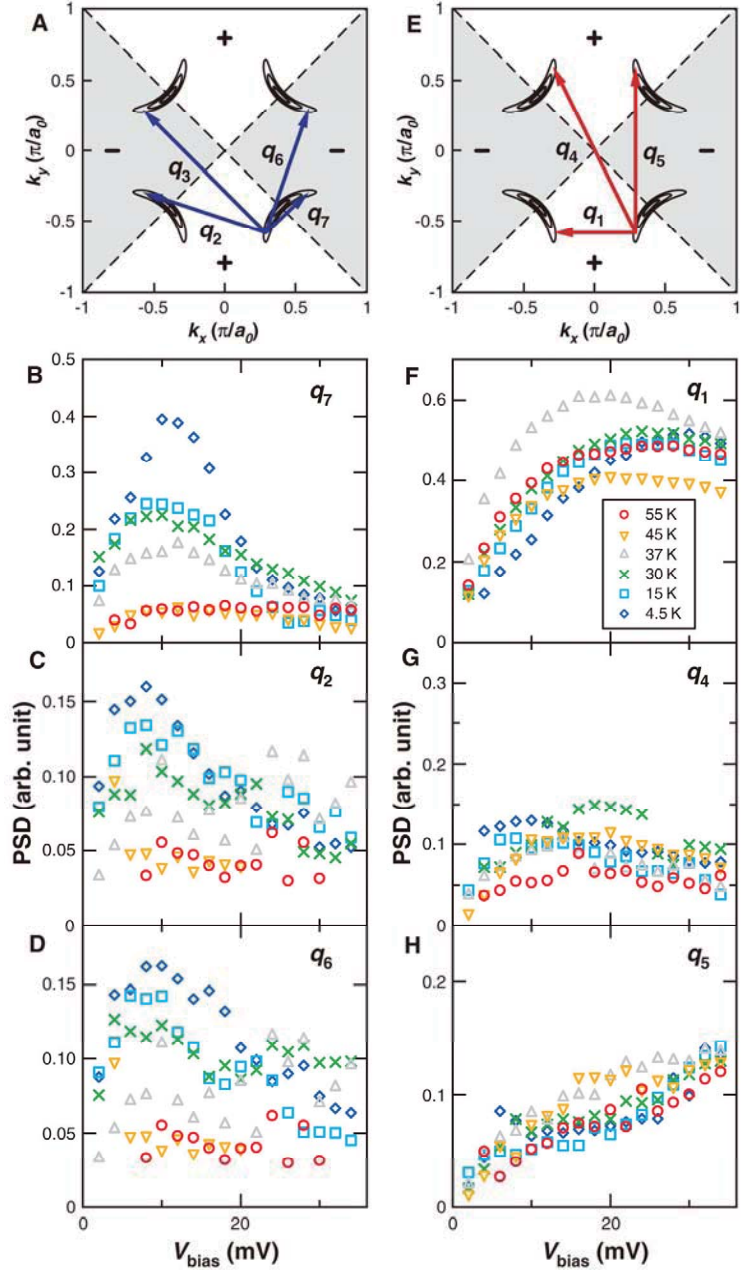

**Figure 4.** (**A**) The octet $q$-vectors due to scattering between regions of $k$-space with different signs of a d-wave order parameter are indicated. (**B** to **D**) The power spectral densities for modulations with $q_7$, $q_2$, and $q_6$ at each energy and for different temperatures are shown. These modulations show a rapid decrease in intensity as temperature is increased. (**E**) The octet $q$-vectors due to scattering between regions of $k$-space with the same sign of a d-wave order parameter are indicated. (**F** to **H**) The power spectral densities for modulations $q_1$, $q_4$, and $q_5$ at each energy and for different temperatures are plotted. The intensity of these modulations shows no trend as temperature is increased and no particular indication of where the transport $T_c$ occurs.

[35] We acknowledge and thank J. C. Campuzano, T. Hanaguri, M. Franz, P. J. Hirschfeld, D.-H. Lee, S. Kivelson, Y. Kohsaka, E.-A. Kim, M. Lawler, J. Lee, K. Levin, D. Morr, M. Norman, N. P. Ong, J. Orenstein, M. Randeria, S. Sachdev, H. Takagi, and A. Tsvelik for helpful discussions and communications. These studies are supported by Brookhaven National Laboratory, the U.S. Department of Energy, the U.S. Office of Naval Research, by Grant-in-Aid for Scientific Research from the Ministry of Science and Education (Japan), and by the Global Centers of Excellence Program for Japan Society for the Promotion of Science. A. S. acknowledges support from the U.S. Army Research Office.

# Supporting Online Material

# Spectroscopic Fingerprint of Phase-Incoherent Superconductivity in the Cuprate Pseudogap State

Jhinhwan Lee, K. Fujita, A. R. Schmidt, Chung Koo Kim, H. Eisaki, S. Uchida, & J.C. Davis*

*To whom correspondence should be addressed. E-mail: jcdavis@ccmr.cornell.edu

## Supporting Text and Figures

**(a) Phase Diagram and the Bogoliubov Quasiparticle Dispersion of Hole-doped Cuprates**

A schematic phase diagram for hole-doped cuprates is shown in Fig. S1A. The regions of the antiferromagnetic insulator (AFI), the pseudogap (PG), and the d-wave superconductor (dSC) are shown. The pale blue region is approximately where the phase-incoherent d-wave superconductivity is detected by thermodynamic and transport experiments. The arrow indicates the temperature range where our experiment was carried out and the hole-density of the sample $Bi_2Sr_2Ca_{0.8}Dy_{0.2}Cu_2O_{8+\delta}$ used. Inset is the magnetic susceptibility of the sample indicating that $T_c$ = 37±3 K.

Figure S1B shows the energy gap and excitation structure in $k$-space for the superconducting phase. The regions of the eight maxima of the density of states at the tips of the $E(\vec{k})$ 'bananas' are connected by the particle-hole symmetric sets of scattering vectors $\vec{q}_i(+E) = \vec{q}_i(-E)$ ($i$=1,2,…,7). These are the 'octet' quasiparticle interference wavevectors appearing in the modulations of $g(\vec{r}, E) = dI/dV(\vec{r}, E = eV)$.

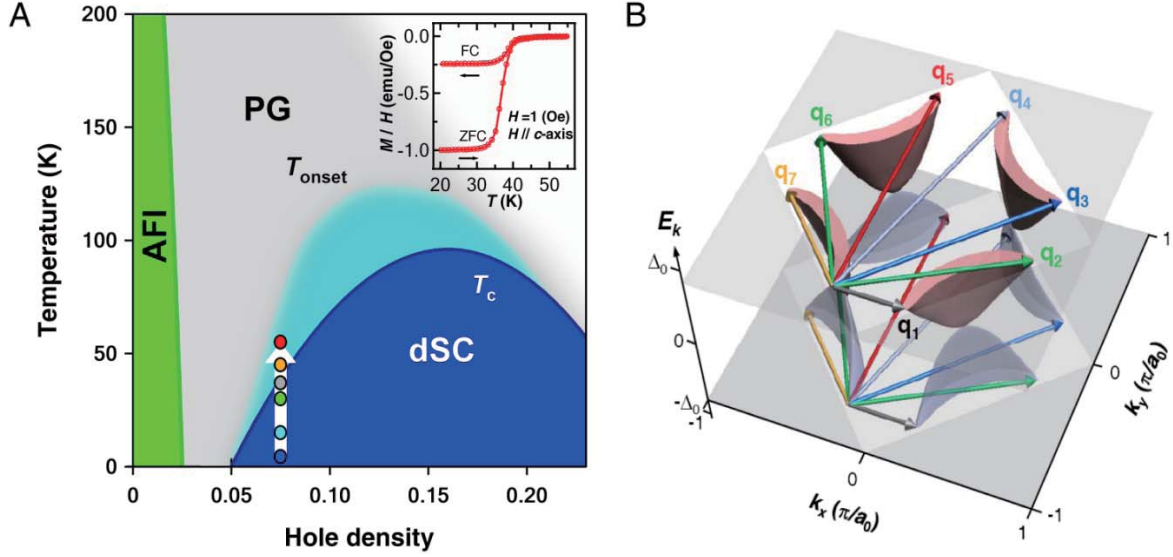

**Figure S1.** (A) The schematic phase diagram of hole-doped cuprate. Temperature range explored in this study with fixed hole-density is indicated by the arrow. (B) The schematic energy gap and the excitation structure in the superconducting phase.

**(b) Experimental Procedures**

We used a specially designed variable temperature (4.4 K ~ 60+ K) spectroscopic imaging STM system, with high (≤1 mK) thermal stability and temperature precision. The combination of the special low power consumption head design and the high Helium-dewar capacity allowed us to take each $dI/dV(\vec{r}, E)$ map for up to 10 days. The cryogenic ultra high vacuum condition enabled measurement of $dI/dV(\vec{r}, E)$ on this single sample surface for more than a year within this temperature range without any detectable surface degradation. We maintained the identical sample and tip conditions and the identical experimental parameters (spectroscopic setpoint voltage 200 mV and setpoint current 200 pA, bias modulation frequency 857.5 Hz, bias modulation amplitude 3 mV$_{rms}$, lock-in time constant 10 ms, averaging time 113 ms, energy range of measurement −34 mV ~ +34 mV and number of discrete energy layers 35) over the whole period of the experiment to extract the true temperature dependence. For this experiment, we developed a special procedure for reproducibly preparing the tungsten tip with identical apex orbital and a flat density of states within 200 mV of its Fermi energy.

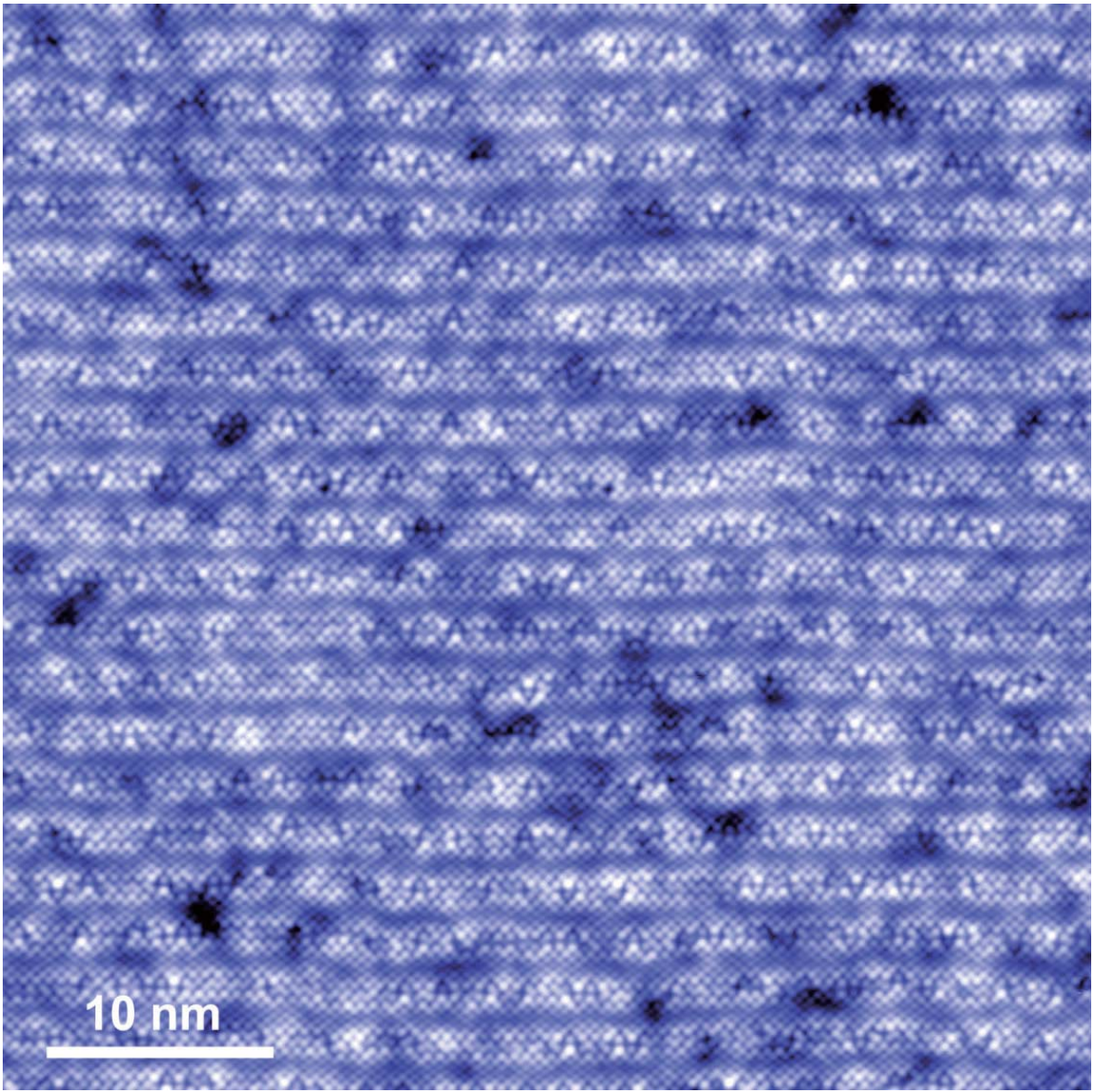

**Figure S2.** 48 nm-wide square field of view of the underdoped ($T_c$ = 37±3 K) $Bi_2Sr_2Ca_{0.8}Dy_{0.2}Cu_2O_{8+\delta}$ sample used for all the QPI maps in this paper. The topograph, taken at 45 K with setpoint voltage 200 mV and setpoint current 20 pA, shows the sample and tip conditions used for all the measurements described in this report.

**(c) QPI Analysis Method**

Throughout this paper we used the ratio

$$Z(\vec{r},E) \equiv \frac{dI/dV(\vec{r},+E)}{dI/dV(\vec{r},-E)} = \frac{N(\vec{r},+E)}{N(\vec{r},-E)} \tag{S1}$$

and its Fourier transform $Z(\vec{q},E)$ to cancel out the projection of the true non-dispersive modulations at high energy onto the low energy $dI/dV(\vec{r},E)$ by the systematic junction formation error. This also selects the particle-hole symmetric QPI modulations. However, due to the inherent nonlinearity in Eq. S1, when any strong modulations (such as $\vec{q}_1$ or $\vec{q}_{atom}=(2\pi/a_0,0)$) exist in the denominator $N(\vec{r},-E)$, they can be mixed in the second order Taylor-expanded term to produce a new $\vec{q}$ component (for example $\vec{q} = \vec{q}_{atom} - \vec{q}_1$) in $Z(\vec{r},E)$ with intensity comparable to the weakest octet peaks. We found that only the $q_5$ peak is appreciably affected by the $\vec{q} = \vec{q}_{atom} - \vec{q}_1$ mixed component at high temperatures and therefore we excluded $q_5$ at all six temperatures from the arc analysis shown in Fig. 3 for overall consistency. Except for the exclusion of $q_5$ in the gap function fitting and the arc analysis, we used identical octet analysis technique used in Ref. *S2*.

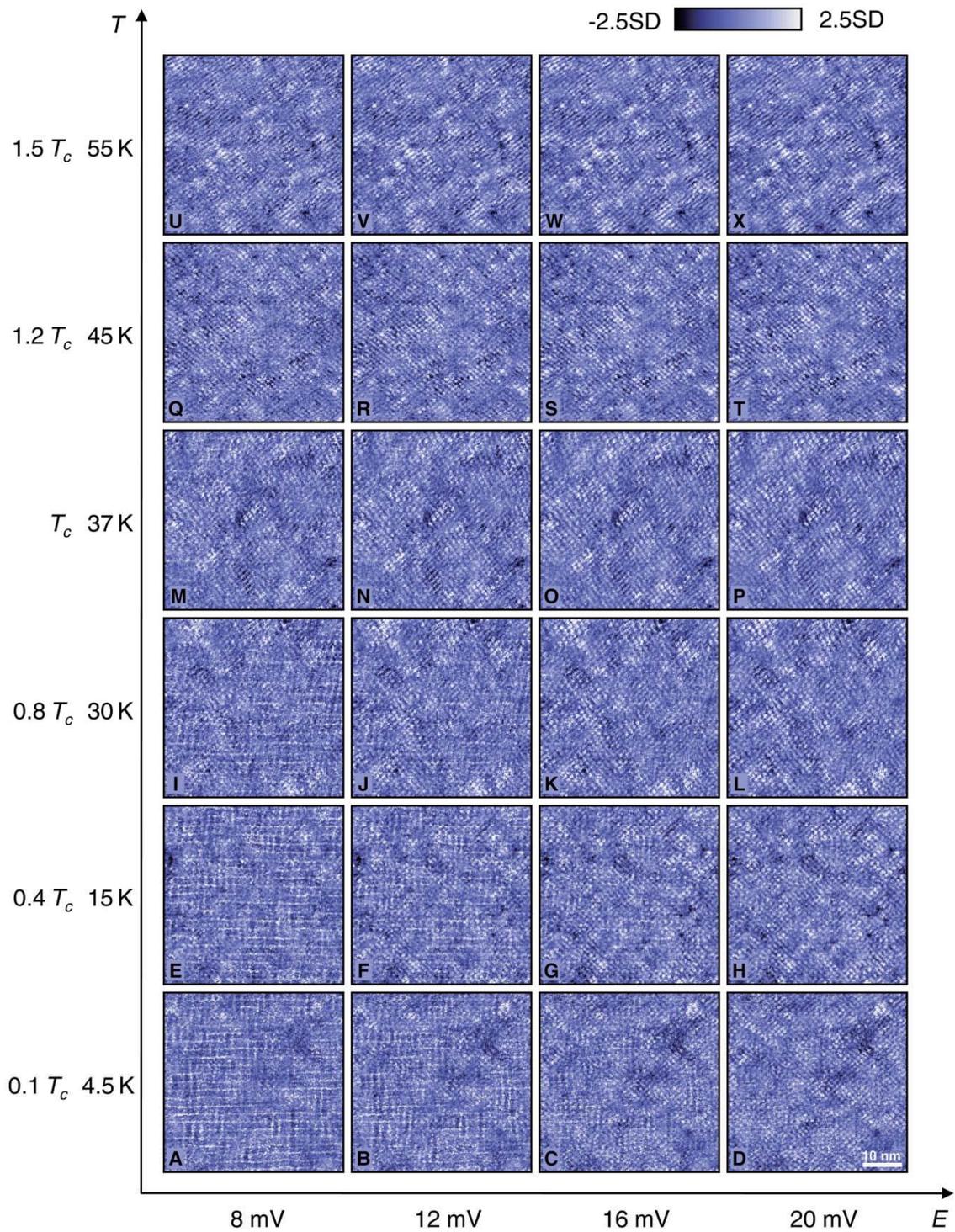

**Figure S3.** A sequence of 45nm-wide unprocessed $Z(\vec{r}, E)$ images as a function of energy and temperature. This is the real-space counterpart of Fig. 1 and the same representative energies and temperatures are used.

Note that in Fig. S3 at low energy and low temperature, the horizontal & vertical lattice-like patterns due to the $q_7$ component are evident. At other energy and temperature conditions the $q_1$ component largely dominates the real space patterns, but all the octet vectors are actually present in all images.

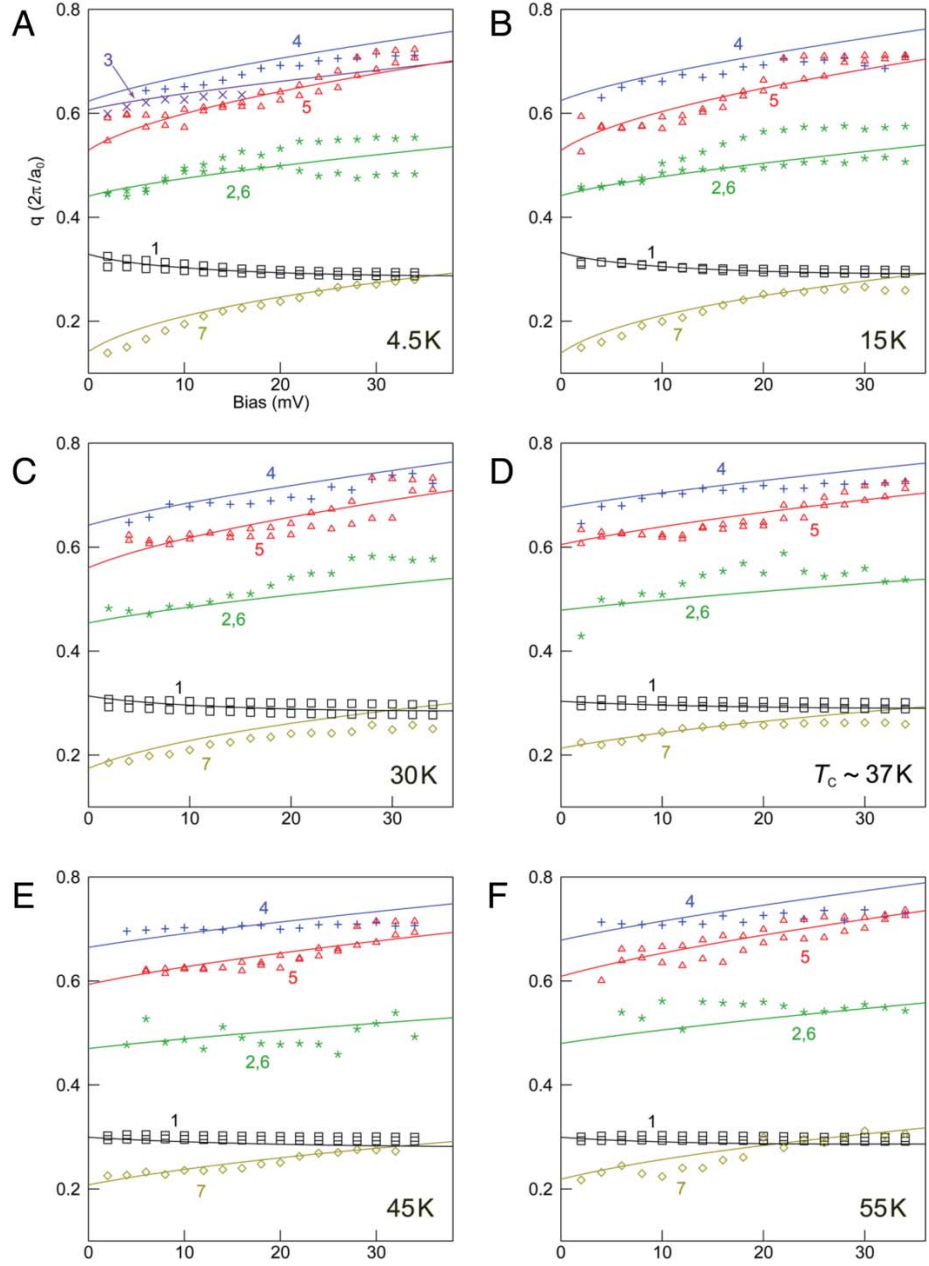

**Figure S4.** The octet model consistency plots at the six temperatures. Note that, at every temperature, all the $q$ vectors conform well to the octet model within their error bars and they disperse in the same way in the superconducting state (A-C) and in the pseudogap state (E-F) without showing any particular indication of $T_c$.

All curves in each panel in Fig. S4 are generated from the combination of the model d-wave gap function $A(T)\cos(2\theta_k) + B(T)\cos(6\theta_k)$ and the normal state Fermi surface modeled with a quarter circle, both fitted using all the available octet $q$ vectors at each temperature. In each panel $q_1$ ($q_5$) is measured in two different equivalent directions (($2\pi/a_0$,0) and (0,$2\pi/a_0$)) and the results are plotted with separate black (red) markers.

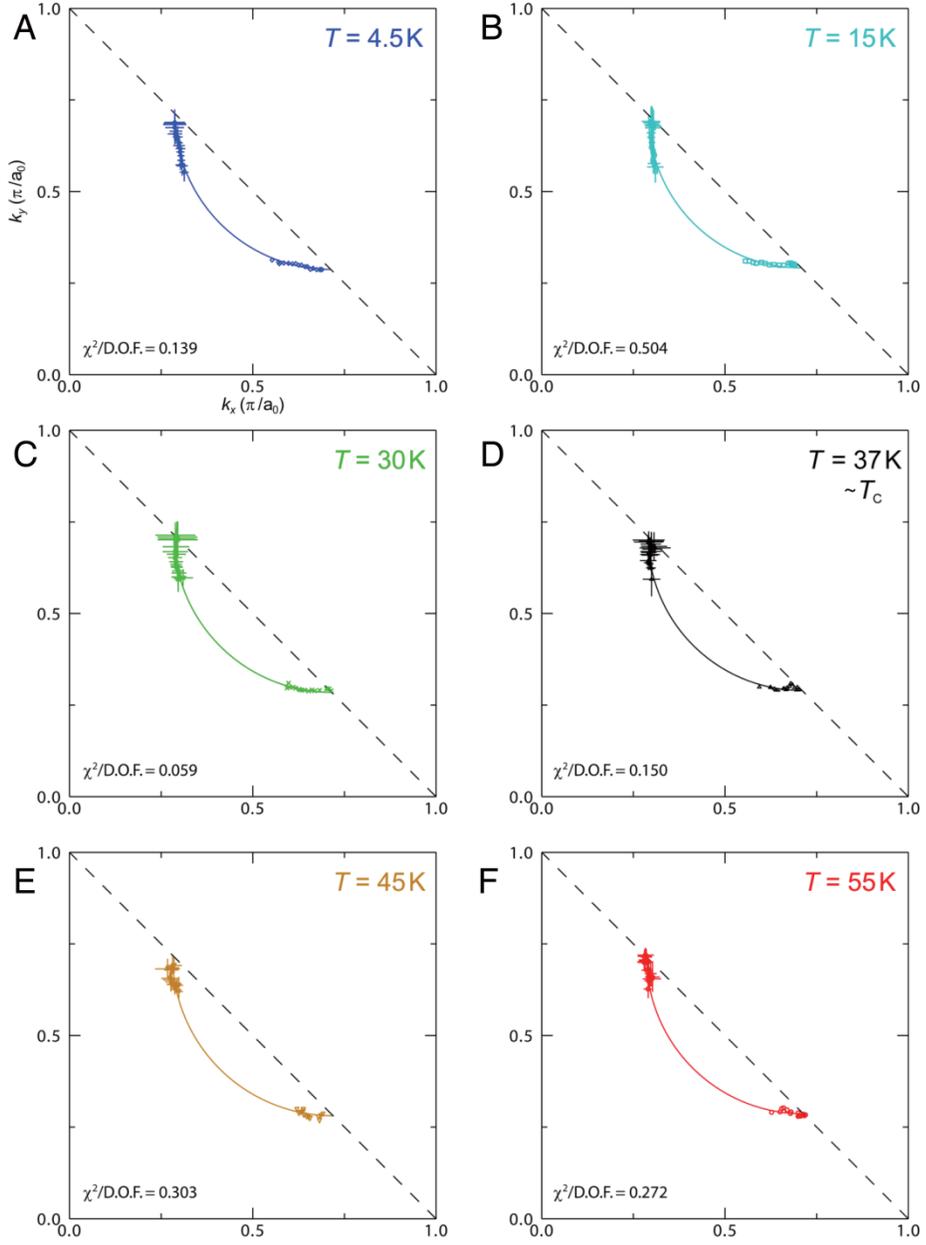

**Figure S5.** The loci of scattering $\vec{k}_B(E)$ (2 mV ≤ |E| ≤ 34 mV) measured at the six temperatures. Note that the d-wave-gapped *k*-space region populated with the data points shrinks and the zero-gap arc near the node expands as the temperature increases from $0.1T_c$ up to $1.5T_c$ without showing any particular sign of $T_c$.

In Fig. S5, a quarter-circle is used as a model of the normal state Fermi surface to fit all the $\vec{k}_B$ points at each temperature. The radius of the fitted quarter circle remained the same at all temperatures within its fitted error bar.

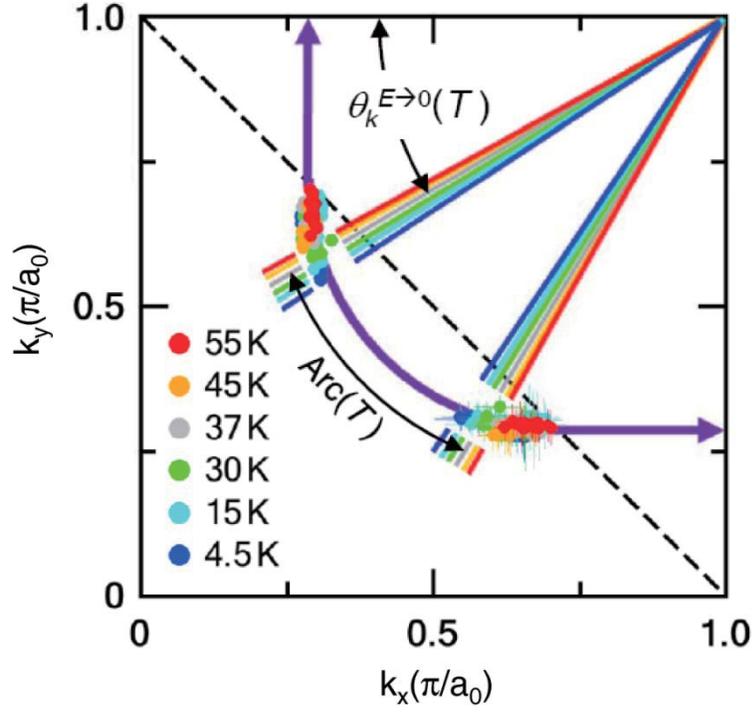

**Figure S6.** The *k*-space loci of scattering at all temperatures.

In Fig. S6, the *k*-space loci of scattering from the internally consistent octet analysis of all the QPI peak locations at all temperatures are shown with solid circles. As temperature is increased, the locus of QPI scattering shrinks along the same *k*-space contour toward the line connecting $(\pi,0)$ and $(0,\pi)$ where the Bogoliubov quasiparticle interference signatures are known to disappear even at lowest temperatures. The locus of QPI scattering remains finite well above $T_c$ (even up to $T = 55$ K) but is confined in a narrower *k*-space region.

**(d) Invariance of High Energy Real Space Electronic Structure above $T_c$**

The real-space R-map patterns which show the local non-dispersive electronic structure at the pseudogap energy scale are also invariant as we increase the temperature well above $T_c$ into the pseudogap state. As shown in Fig. S7, in the pseudogap state ($T$ = 45 K > $T_c$) we continue to observe the characteristic R-map patterns of broken translational & rotational electronic symmetry and bond-centered peaks identical to those seen at low temperature (4 K) and at the pseudogap energy scale in underdoped Bi2212 and Na-CCOC [*S3*].

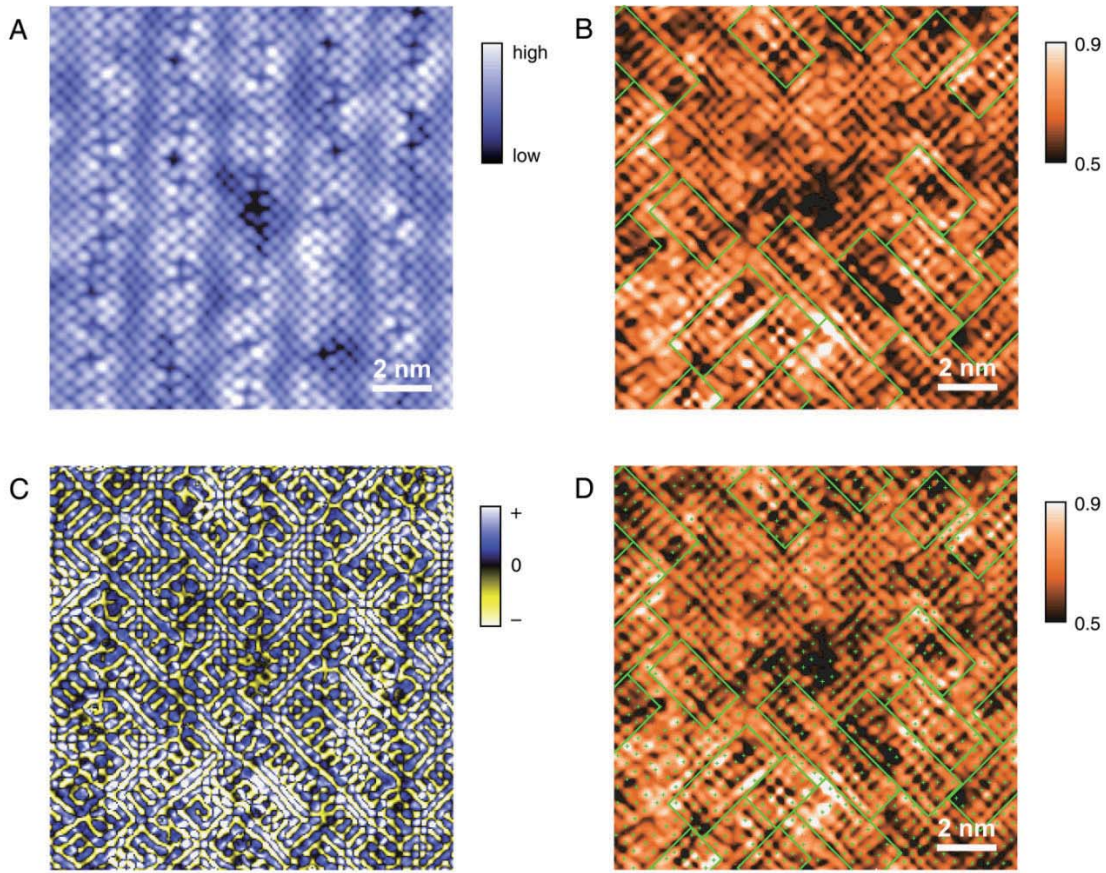

**Figure S7.** 13.5 nm-wide topograph (A) and R map ( $R(\vec{r}) \equiv I(\vec{r},+200\,\text{mV})/I(\vec{r},-200\,\text{mV})$ ) (B-D) simultaneously taken at $T$ = 45 K > $T_c$ with setpoint voltage 200 mV and setpoint current 100 pA. In C, Laplacian filtering is used to enhance the contrast of the atomic scale R-map patterns. In B and D, the green rectangles are used as guides to the eyes for the characteristic R-map patterns with locally broken translational and rotational symmetry and bond-centered peaks. In D, the top-layer Bi atom positions (which approximately correspond to the underlying Cu atom positions) in the topograph (A) are marked with green crosses.

## (e) Arc Length Analysis

In order to precisely determine the zero-gap arc length without relying on any specific d-wave gap function model, we used linear regression scheme applied only to the low energy data points as explained below.

The low energy ($|E_i| \leq 14$ mV) data points $\theta_k(E_i) \pm \delta\theta_k(E_i)$ at each temperature are fitted to a linear function $\theta_k = \alpha + \beta \cdot E$ with weights given by their error bars $\delta\theta_k(E_i)$ as shown in Fig. S8A, where we obtain the expectation values of $\alpha$ & $\beta$ and their errors $\delta\alpha$ & $\delta\beta$ (with 95% confidence). The intercepts of the fitted lines with the $E = 0$ axis ($\alpha \pm \delta\alpha$, marked with triangles) correspond to the predicted $\theta_k^{E\to 0} \pm \delta\theta_k^{E\to 0}$ that are later converted to the arc lengths shown with the solid squares with error bars in Fig. 3B.

The conversion from the arc angle $\theta_k^{E\to 0}$ (Fig. S8B) to the arc length (%) (Fig. 3B) is done based on the normal state Fermi surface (the purple curved arrow in Fig. S6) modeled by a quarter circle inside the first $\sqrt{2}\times\sqrt{2}$ Brillouin zone and two hypothetical anti-nodal straight lines running orthogonal to the 1x1 Brillouin zone boundary as shown in Fig. S6. The arc length (%) in Fig. 3B is defined as the % ratio of the length of $Arc(T)$ to the total length of the purple curve in Fig. S6.

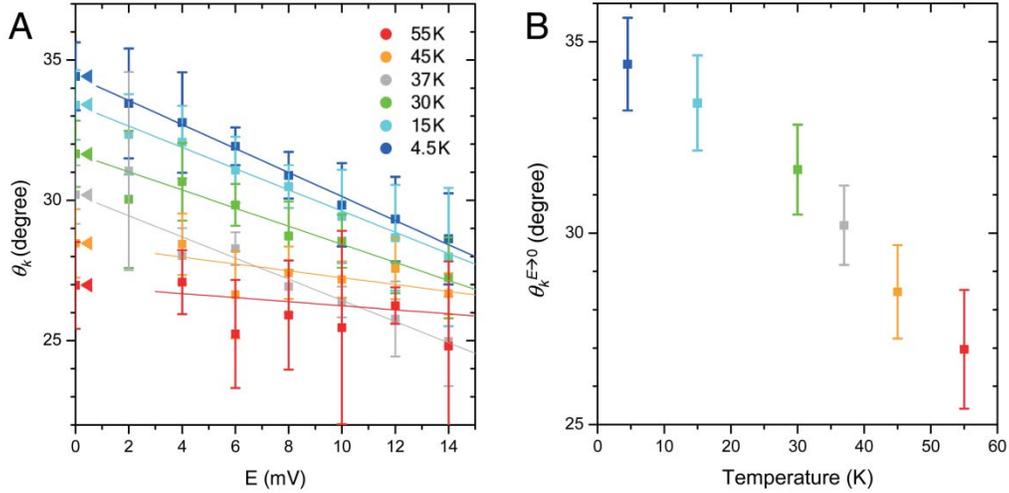

**Figure S8.** (A) Low energy linear regression scheme used in determination of the zero gap arc angle and its error ($\theta_k^{E\to 0} \pm \delta\theta_k^{E\to 0}$). The intercepts of the fitted lines with the $E = 0$ axis, marked with triangles, correspond to the predicted $\theta_k^{E\to 0}$. (B) Plot of zero gap arc angle vs. temperature ($0.1T_c \sim 1.5T_c$). Note the monotonicity as a function of temperature even though the temperature-dependence of the fitted slopes shown in A is not strictly monotonic.